\documentclass{pasa}%

\title[Dynamical gravitational wave source]{A dynamical gravitational wave source in a dense cluster}
\author[Hurley et al.]{Jarrod R. Hurley$^{1,3}$\thanks{jhurley@swin.edu.au}, Anna C. Sippel$^2$, Christopher A. Tout$^3$ \and Sverre J. Aarseth$^{3}$\\
\affil{$^1$Centre for Astrophysics and Supercomputing, Swinburne University of Technology, P.O. Box 218, VIC 3122, Australia}%
\affil{$^2$Max Planck Institute for Astronomy, K{\"o}nigstuhl 17, 69117 Heidelberg, Germany}%
\affil{$^3$Institute of Astronomy, University of Cambridge, Madingley Road, Cambridge, CB3 0HA, UK}}%
\jid{PASA}
\doi{10.1017/pas.\the\year.xxx}
\jyear{\the\year}

\usepackage[authoryear]{natbib}
\bibpunct{(}{)}{;}{a}{}{,}
\usepackage{aas_macros}
\usepackage{hyperref} 
\hypersetup{colorlinks,citecolor=blue,linkcolor=blue,urlcolor=blue}

\usepackage{tikz}
\usepackage{pgfplots}

\begin{document}%
\begin{abstract}
Making use of a new $N$-body model to describe the evolution of a moderate-size globular cluster 
we investigate the characteristics of the population of black holes within such a cluster. 
This model reaches core-collapse and achieves a peak central density typical of the 
dense globular clusters of the Milky Way. 
Within this high-density environment we see direct confirmation of the merging of two 
stellar remnant black-holes in a dynamically-formed binary, a gravitational wave source. 
We describe how the formation, evolution and ultimate ejection/destruction of binary systems 
containing black holes impacts the evolution of the cluster core. 
Also, through comparison with previous models of lower density, we show that the 
period distribution of black hole binaries formed through dynamical interactions in 
this high-density model 
favours the production of gravitational wave sources. 
We confirm that the number of black holes remaining in a star cluster at late 
times and the characteristics of the binary black hole population depend on the 
nature of the star cluster, critically on the number density of stars and by extension the relaxation 
timescale. 
\end{abstract}
\maketitle%
\section{INTRODUCTION }
\label{sec:intro}

The recent detection of gravitational waves (Abbott et al. 2016a) 
believed to be from the merging of two black holes (BHs) has invigorated the 
modelling community and led to a new set of papers on BH--BH merging rates 
expected from star cluster populations 
(Chatterjee, Rodriguez \& Rasio 2016; Mapelli 2016; Rodriguez, Chatterjee \& Rasio 2016; Rodriguez et al. 2016) 
and field binaries 
(Belczynski et al. 2016a; Eldridge \& Stanway 2016). 
These studies build on work over the past decade or more that focussed on predictions for the 
Laser Interferometric Gravitational-wave Observatory (LIGO) and Advanced LIGO detections 
of the merging of double compact object binaries with population synthesis of binary stars (e.g. Belczynski, Kalogera \& Bulik 2002; Dominik et al. 2015) 
and models of the dense star cluster environment 
(e.g. Portegies Zwart \& McMillan 2000; Downing et al. 2011; Antonini et al. 2016). 

The gravitational wave (GW) source GW150914 is the first detection of its kind and also the first observational 
evidence for the merging of a BH--BH binary (Abbott et al. 2016a). 
The component masses are derived to be $36$ and $29 \, M_\odot$ based on signal matching 
the waveform expected for the inspiral and merge of these BH masses. 
 It has been suggested that a globular cluster (GC) environment is the most likely place to 
 form such a high-mass BH binary (Rodriguez et al. 2016) 
but pathways exist for normal binary evolution in the field as well (Belczynski et al. 2016a; 
see also the discussion in Dvorkin et al. 2016). 
More recently a second source GW151226 has been announced (Abbott et al. 2016b) 
with derived masses of $14.2$ and $7.5 \, M_\odot$ (although with sizeable error bars). 
Previous studies have examined how a population of BHs may evolve in a star cluster 
(see Breen \& Heggie 2013 and Chatterjee, Rodriguez \& Rasio 2016 for recent summaries). 
In essence, many BHs are expected to form in a typical GC and these BHs  
quickly form a centralised subsystem which may be unstable (Spitzer 1969) owing to 
the formation of BH binaries and the ejection of single BHs as well as the binaries in strong 
encounters (Kulkarni, Hut \& McMillan 1993; Sigurdsson \& Hernquist 1993). 
However, numerical studies of star cluster evolution have shown that expansion of the BH 
subsystem can occur as a result of these ejections, increasing the timescale for BH 
evaporation and suggesting that a sizeable population of BHs can reside in present day GCs 
(e.g. Breen \& Heggie 2013; Morscher et al. 2015). 
Portegies Zwart \& McMillan (2000) postulated that the BH binaries that form in the 
centralised BH subsystem of a star cluster are ripe for merging and thus candidates 
for detection as gravitational waves. 
The occurrence of merges on short gravitational radiation timescales within the 
dense stellar environment of a cluster has been confirmed via $N$-body simulation by Aarseth (2012). 
It has also been shown that these BH binaries can in turn influence the overall 
evolution of the host cluster (Hurley \& Shara 2012; Morscher et al. 2015). 

Here we introduce an $N$-body model of $N = 2 \times 10^5$ which reaches a high 
central density (about $10^6 \,$stars ${\rm pc}^{-3}$) and the end of the core-collapse 
phase at about $12\,$Gyr. 
This complements recent higher-$N$ models such as that of Wang et al. (2016)  
which did not reach core-collapse and previous lower density models of 
$N \simeq 2 \times 10^5$ (Sippel \& Hurley 2013). 
It can also be compared to Monte Carlo models used to investigate the formation and behaviour 
of BH--BH binaries in star clusters (Morscher et al. 2015; Rodriguez et al. 2016). 
We describe the setup of the model and general evolution characteristics in Section~2 
then look specifically at the BH population within the model cluster in Section~3. 
In Section~4 we highlight a dynamically influenced BH--BH merger which occurred within 
the cluster, describing the formation pathway and outcome, then look at the effect of BH binary 
merges and ejection events on the behaviour of the cluster core. 
Finally we compare the period distribution of dynamically formed BH--BH binaries in 
high- and low-density cluster models (Section~5) and discuss the results.

\section{THE HIGH DENSITY $N$-BODY STAR CLUSTER}

We focus here on a simulation that started with $195\,000$ single stars and $5\,000$ primordial 
binaries and was evolved to an age of $16\,$Gyr using the direct $N$-body code 
{\tt NBODY6} (Aarseth 2003; Nitadori \& Aarseth 2012). 
The code includes algorithms to follow single and binary star evolution, as described by  
Hurley et al. (2001), in step with the calculation of the gravitational forces which are 
integrated with a 4th-order Hermite scheme. 
Particular attention is paid to the treatment of close encounters, with regularization schemes 
and stability algorithms employed to deal efficiently with small-$N$ subsystems 
(see Aarseth 2003 for details), while collisions, merges, dynamical perturbations of binary 
orbits, and exchange interactions, for example, are allowed. 

\begin{figure*}
\begin{center}
\includegraphics[width=\columnwidth]{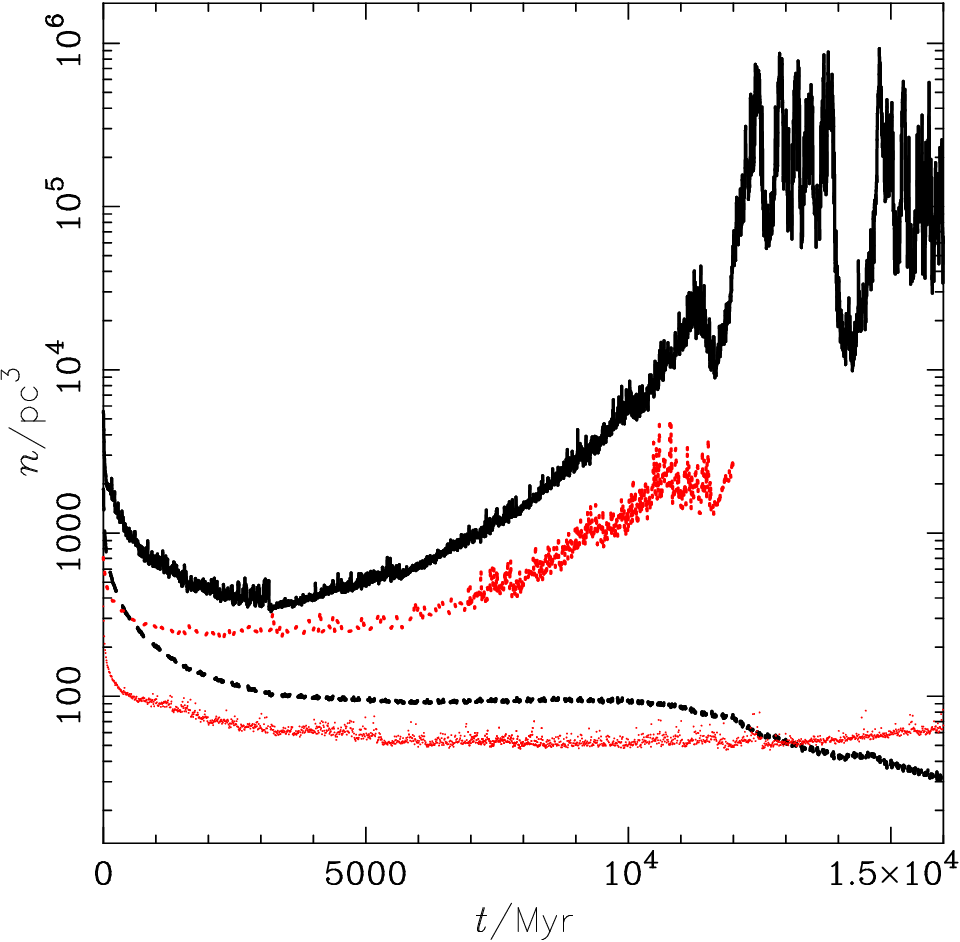}
\caption{The stellar number density in the cluster core as a function of age for the new model 
presented in this work (black sold line), the Hurley \& Shara (2012) model (red dotted line) 
and the Sippel \& Hurley (2013) model (red points). 
Also shown for reference is the number density within the half-mass radius for the new model (dashed line).}
\label{Fig1}
\end{center}
\end{figure*}

The stellar masses are chosen from a Kroupa (2001) initial mass function (IMF) between 
the limits of $0.1$ and $50\, M_\odot$. 
For the binary stars we combine the chosen masses to give the binary mass 
and then redistribute the component masses according to a mass-ratio drawn at random 
from a uniform distribution. 
Orbital parameters, period and eccentricity, are chosen according to the 
method described by Geller, Hurley \& Mathieu (2013) to match the observed characteristics of the 
binary population in the young open cluster M35. 
Specifically this means that orbital periods follow the Duquennoy \& Mayor (1991) 
log-normal period distribution and eccentricities follow a Gaussian distribution centred 
on $e = 0.38$ with $\sigma = 0.23$. 
The metallicity chosen for the stars is $Z = 0.0002$ (or $\left[ {\rm Fe} / {\rm H} \right] \simeq -2$). 
This is the metallicity of the globular cluster NGC$\,6397$ which is part of the metal-poor sample 
of Milky Way globular clusters 
(approximately 10 per cent of the sample have lower metallicities: Bica et al. 2006). 
At this metallicity stars with initial masses $18.4 \,  M_\odot$ or greater evolve to become 
BHs while stars of $6.1 \leq M /  M_\odot < 18.4$ become neutron stars (NSs). 
The main-sequence turn-off mass at an age of $12\,$Gyr is $0.83 \, M_\odot$ and stars with 
$0.84 < M / M_\odot < 6.1$ have evolved to become white dwarfs (WDs) by this age. 
All stars are assumed to be on the zero-age main sequence when the simulation begins. 
Stellar evolution is supplied by the Single Star Evolution (SSE) algorithm 
described by Hurley, Pols \& Tout (2000). 
We use the SSE prescription for mass loss in stellar winds except that, 
following the discussion by Belczynski et al. (2010), the luminous blue variable 
mass-loss rate for massive stars has not been applied because it overly restricts BH masses 
for low- to intermediate-metallicity populations. 
Another difference is that NS and BH remnant masses are set following the updated procedure 
of Belczynski, Kalogera \& Bulik (2002).

It is generally assumed that NSs and possibly BHs receive a velocity kick at birth owing 
to asymmetries in the preceding core-collapse supernovae. 
The magnitude of these kicks and how they are affected by complications such as the fallback 
of material during BH formation are uncertain. 
Some guidance can be gleaned from the observed space velocities of pulsars which 
are typically of the order of hundreds of ${\rm km} \, {\rm s}^{-1}$ which, when compared to 
the typical escape velocity of a star cluster, of order $10 \, {\rm km} \, {\rm s}^{-1}$ or less,  
leads to a problem of retaining supernova remnants (Pfahl, Rappaport \& Podsiadlowski 2002). 
While there are models that suggest a mass-dependent kick distribution for BHs 
(Belczynski, Kalogera \& Bulik 2002) this is tempered by the results of 
Repetto, Davies \& Sigurdsson (2012) who find that BHs and NSs require similar kicks 
in order to explain the observed distribution of low-mass X-ray binaries in the Milky Way. 
For simplicity we assign NSs and BHs a velocity kick at birth chosen at random from a uniform 
distribution between $0$ and $100 \, {\rm km} \, {\rm s}^{-1}$. 
The choice of natal kick distribution will have implications for the nature of the BH-BH 
population and this will be discussed in Section~5.

\begin{figure*}
\begin{center}
\includegraphics[width=\columnwidth]{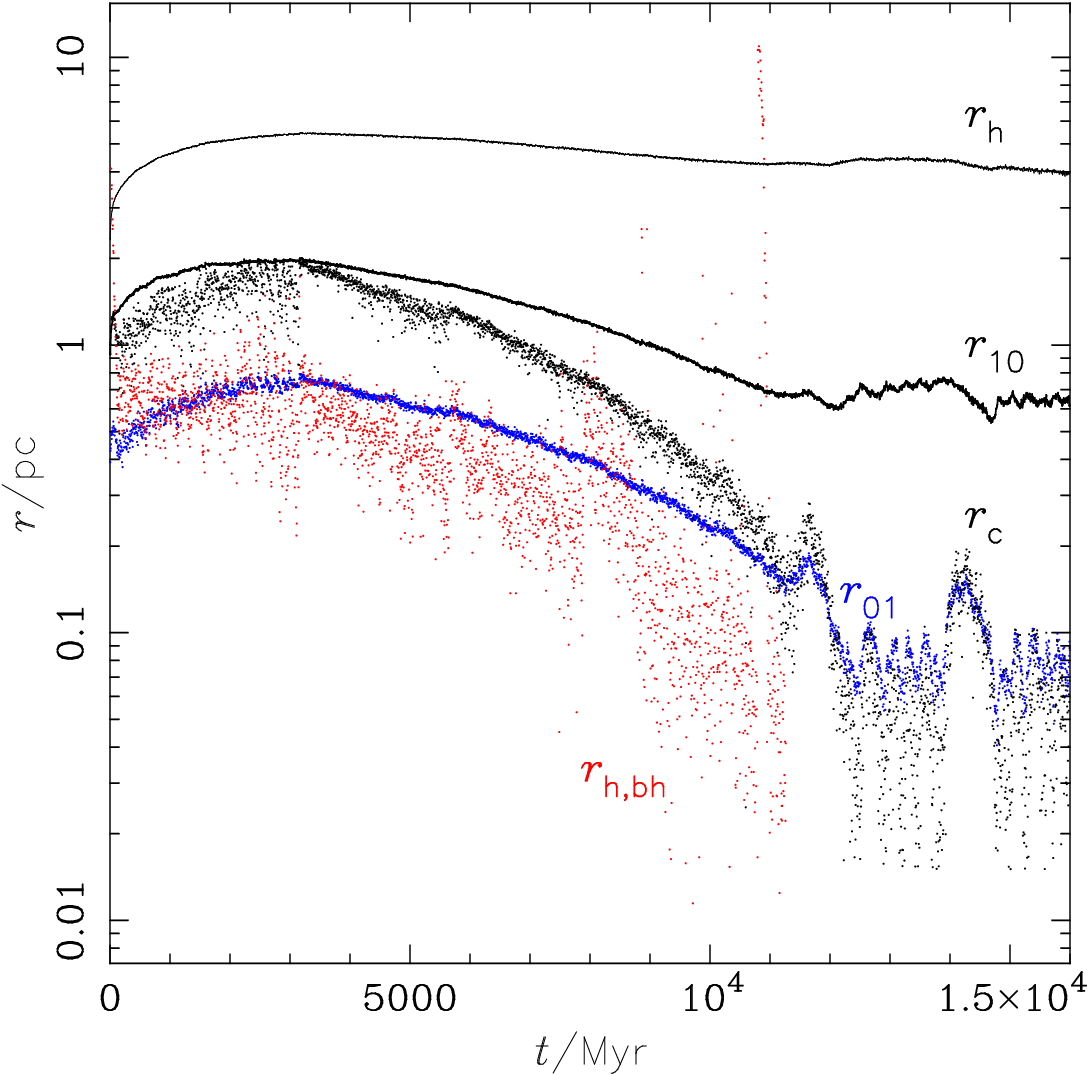}
\caption{Evolution of various cluster radii with age: half-mass radius ($r_{\rm h}$, upper black line), 
inner Lagrangian 10 per cent radius ($r_{10}$, lower black line), core radius ($r_{\rm c}$, black points), 
inner Lagrangian 1 per cent radius ($r_{01}$, blue points), and the half-mass radius of 
the black holes ($r_{\rm h,bh}$, red points).
The black hole half-mass radius is only plotted when more than two BHs are in the cluster. }
\label{Fig2}
\end{center}
\end{figure*}

We assign the initial positions and velocities of the cluster members according to a Plummer 
density profile (Plummer 1911; Aarseth, H\'{e}non \& Wielen 1974) and put the stars 
and binary systems initially in virial equilibrium. 
The half-mass radius of this starting model is $2.3 \,$pc and the outermost 
star is $18.1 \,$pc from the cluster centre. 

The host galaxy is modelled as a three-component Milky Way consisting of a 
point-mass bulge of $1.5 \, \times \, 10^{10} \, M_\odot$ (Xue et al. 2008), 
a Miyamoto \& Nagai (1975) disc of $5 \, \times \, 10^{10} \, M_\odot$ 
(Xue et al. 2008) and scale-lengths $a = 4\,$kpc, $b = 0.5\,$kpc, 
and a logarithmic dark matter halo set such that the combined mass of the bulge, disc 
and halo gives a circular velocity of $220 \, {\rm km} \, {\rm s}^{-1}$ at a distance of 
$8.5\,$kpc from the Galactic centre. 
The sum of the contributions from the force derivatives of the three components 
is included directly in the force calculations, as described in Aarseth (2003). 
Within this host galaxy the cluster is placed at an apogalacticon of 
$\left( x, y, z \right) = \left( 7, 0, 0 \right) \,$kpc 
with a velocity vector $\left( 0, 115, 45 \right) \, {\rm km} \, {\rm s}^{-1}$. 
So the cluster orbits between a radial distance of $2.7$ to $7.0\,$kpc within 
the disc and extends to about $0.9\,$kpc above and below the plane of the disc. 
The initial mass of the cluster is $1.26 \times 10^5 \, M_\odot$. 
This gives a tidal radius of approximately $50\,$pc at apogalacticon. 
Accordingly we set an escape radius of $100\,$pc beyond which we consider 
stars (or binaries) to no longer be cluster members.

After $12\,$Gyr of evolution the cluster mass, that of bound members, has  
reduced to $3.0 \times 10^4 \, M_\odot$ and at $16\,$Gyr it is $1.2 \times 10^4 \, M_\odot$. 
We note that the orbit of the model cluster is similar to that of the Milky Way globular 
cluster NGC$\,6397$ (Kalirai et al. 2007) which we used as a guide. 
However, the current mass of that cluster is estimated to be anywhere from 
$6 \times 10^4 \, M_\odot$ (Drukier 1995; Giersz \& Heggie 2009) to 
$2.5 \times 10^5 \, M_\odot$ (Pryor \& Meylan 1993), at least a factor of 2 greater 
than this model at a similar age. 

Fig.~\ref{Fig1} describes how the stellar number density of the model evolves 
with cluster age. 
Shown are the number density within the cluster core (upper solid line) and within 
the half-mass radius (dashed line). 
Both are the number of stars divided by the relevant volume. 
Also shown for comparison are the core number density behaviour in the 
$N = 200\,000$ model of Hurley \& Shara (2012) and in the $N = 250\,000$ model 
of Sippel \& Hurley (2013). 
The Hurley \& Shara (2012) model was evolved on a circular orbit at a distance of 
$3.9\,$kpc from the centre of the galaxy. 
It experienced core-collapse at an age of $10.5\,$Gyr and had $1.7 \times 10^4 \, M_\odot$ 
remaining when the simulation ended at $12\,$Gyr. 
That simulation was notable for the ejection of a BH--BH binary at about $11.5\,$Gyr. 
This caused a sharp increase in the size of the core and a corresponding drop in core 
number density, which is evident in Fig.~\ref {Fig1}, erasing the signature of core collapse. 
The Sippel \& Hurley (2013) model was evolved on a circular orbit at a distance of $8.5\,$kpc 
from the centre of the galaxy. 
It had $6.7 \times 10^4 \, M_\odot$ of bound mass in stars and binaries remaining at an 
age of $12\,$Gyr. 
This model did not experience core collapse in the $16\,$Gyr timeframe owing to a 
longer half-mass relaxation timescale compared to the other models. 

It is noticeable from Fig.~\ref{Fig1} that our new model reaches a much higher central 
density than the previous models, 
at least an order of magnitude greater in fact, with the central density of the Sippel \& Hurley (2013)  
model comparable to the half-mass radius density of the new model which in turn can be as much 
as four orders of magnitude less than the central density. 
The peak central density of the new model is about $10^6 \,$stars ${\rm pc}^{-3}$ which 
is approaching that expected for the dense globular clusters of the Milky Way,  
as listed by Pooley et al. (2003). 
Note that luminosity densities 
can be converted to indicative number densities by assuming an average stellar mass of $0.5 M_\odot$ 
and a mass-to-light ratio of $M/L = 2 \, M_\odot / L_\odot$ (Sippel et al. 2012). 

We see a decrease in the central density over the first few Gyr of evolution as the cluster 
expands in response to stellar evolution induced mass loss. 
This is followed by a period of sustained density increase associated with the main 
core-collapse phase of the cluster evolution, which is interrupted briefly by a dip 
in density at about $11.5\,$Gyr (discussed below). 
The core-collapse phase ends at $12.2\,$Gyr with a central density of 
approximately $10^6 \,$stars ${\rm pc}^{-3}$ and is followed by a series of core oscillations as well 
as a more significant drop in density at around $14\,$Gyr (also to be discussed below). 

The corresponding core radius evolution is shown in Fig.~\ref{Fig2}. 
We calculate the $N$-body core radius using the density-weighted procedure 
of Casertano \& Hut (1985). 
As described by Sippel et al. (2012) fluctuations are to be expected, owing to the actions 
of a few massive BHs or energetic binaries in the central regions. 
We see from Fig.~\ref{Fig2} that, for the majority of the evolution, the core radius sits between 
the inner Lagrangian $1\,$per cent and $10\,$per cent radii, that is the radius that contains the inner $1\,$per cent and 
$10\,$per cent of the cluster mass, respectively. 
At late times (about $10\,$Gyr or later) the core radius and the$1\,$per cent Lagrangian radius 
effectively track each other, although with some extended fluctuation in the core radius. 
The half-mass radius expands from its initial $2.3\,$pc over the first $2\,$Gyr 
of evolution and then remains fairly steady within the $4-5\,$pc range for the remainder 
of the evolution. 
We note that the half-mass relaxation timescale (Spitzer 1987) of the initial model was $350\,$Myr, 
increasing to a maximum of about $1\,500\,$Myr at $3.2\,$Gyr and decreasing 
to $600\,$Myr at $12\,$Gyr.

\section{THE BLACK HOLE POPULATION}

Approximately $380$ BHs and $1\,530$ NSs form in the model. 
As mentioned above, these come primarily from stars with initial masses in the ranges 
$18.4 \leq M / M_\odot \leq 50$ and $6.1 \leq M / M_\odot < 18.4$, respectively, with binary evolution 
having the ability to blur the boundaries. 
The first supernova occurs after about $4\,$Myr of evolution. 
At an age of $50\,$Myr, by which time all BHs have formed, there are 70 remaining in the 
cluster, a retention rate of roughly 18 per cent after velocity kicks have been applied. 
NS formation continues until $120\,$Myr and 14 per cent are retained at this age. 

Fig.~\ref{Fig2} shows the evolution of the half-mass radius of the BH subsystem 
which can be compared to the half-mass radius of all stars as well as other key radii. 
The mass range of the BHs is typically $5 < M / M_\odot < 30$ so they quickly become the 
dominant population by mass. 
Notably the BH half-mass radius is generally smaller than the radius containing 
the inner 1 per cent of the mass of all stars, indicating that the BHs are a centrally concentrated 
subpopulation. 
In fact, both the core radius and the inner $1\,$per cent Lagrangian radius exhibit a sharp 
decrease between $100$ and $200\,$Myr, down to about $0.4\,$pc. 
This can be attributed 
to an early collapse of the BH subsystem which has its own much shorter relaxation timescale relative to the general stellar population. 
This behaviour has been noted by other authors in the past 
(Portegies Zwart \& McMillan 2000; 
Chatterjee, Rodriguez \& Rasio 2016; Wang et al. 2016). 

The number of BHs in the cluster steadily decreases as the cluster evolves. 
Of the 70 present at $50\,$Myr there are 43 remaining at $1\,$Gyr, 
20 remaining at $3\,$Gyr and 10 at $7\,$Gyr. 
By the time the cluster has reached an age of $10\,$Gyr only five BHs remain 
and at core-collapse ($12.2\,$Gyr) there is a solitary BH in the cluster. 
This steady decrease of the BH population is similar to that found by Sippel \& Hurley (2013) 
although in their lower density model, with a much longer dynamical timescale, there 
were 16 BHs remaining after $12\,$Gyr of evolution. 
The prime cause of the decrease is the velocity imparted to BHs in few-body 
encounters. 
This results in escape from the cluster. 

The evolution of the half-mass radius of a BH subsystem was studied by Breen \& Heggie (2013) 
using two component $N$-body models where all BHs have mass $m_2$ and all other 
stars have mass $m_1$, with $m_2 > m_1$. 
After the initial collapse phase of the BH subsystem they find that there is a long-lived 
phase of energy balance between this centralised subsystem and the remainder of the 
cluster. 
They demonstrate a relationship between the half-mass radius of the BH subsystem 
($r_{\rm h,bh}$) 
and the overall half-mass radius ($r_{\rm h}$) in terms of the total mass in BHs ($M_2$) 
and the total mass of all other stars ($M_1$) where  
$r_{\rm h,bh} / r_{\rm h} \simeq 0.33 \left( M_2 / M_1 \right)^{0.28}$. 
For our model at $7\,$Gyr, an age which is well separated from the early core-collapse 
of the BH population and the later end of the core collapse phase for all stars, as well 
as still having a sizeable BH population of 10 to ensure reasonable statistics: 
$M_2 / M_1$ is $0.0011$ (BHs have about 0.1 per cent of the cluster mass). 
Placing this into the relationship of Breen \& Heggie (2013) gives a prediction 
of $r_{\rm h,bh} / r_{\rm h} \simeq 0.05$ which is an excellent match 
to the actual value of the model. 
This lends credence to the validity of their relationship and means that we can assume that 
the BH subsystem is in the energy balance phase at this time. 
We note that $m_2 / m_1$, the ratio of the average BH mass to the average stellar mass, 
is about $15$ for our model at $7\,$Gyr compared to $m_2 / m_1 = 10$ used by 
Breen \& Heggie (2013) to develop their relationship. 
For their series of $N = 10^6$ models, Wang et al. (2016) also found consistency 
with the relationship of Breen \& Heggie (2013), even with a higher $m_2 / m_1 \simeq 40$. 

\begin{figure*}
\begin{center}
\includegraphics[width=15cm]{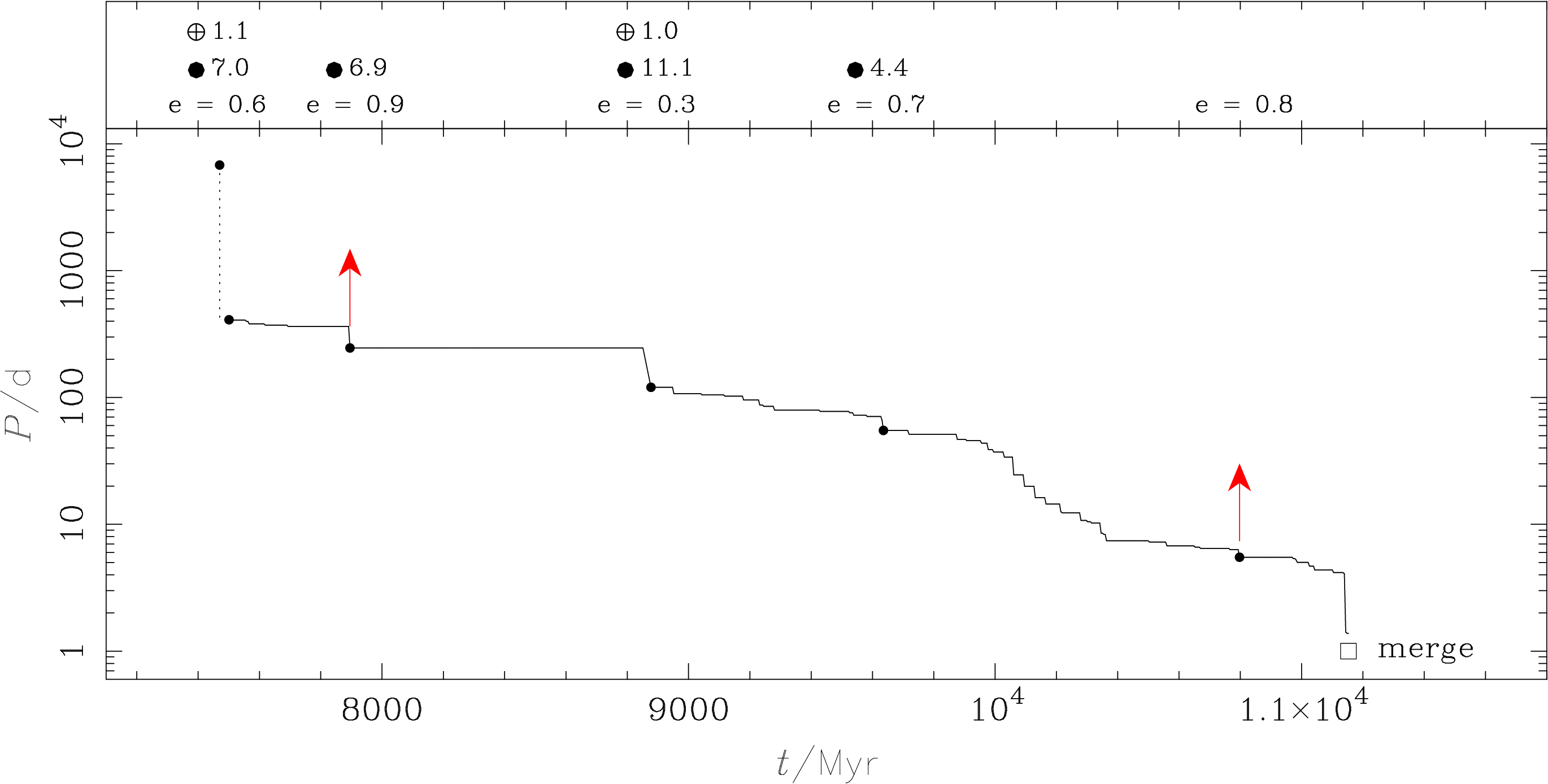}
\caption{Period evolution of the BH--BH binary comprising $9.1$ and $8.2 \, M_\odot$ BHs that ends by merging. 
Some of the major interactions involving this system are marked along the period 
evolution path (corresponding to the descriptions in the text). 
The single stars involved in these interactions are listed above each marked point:  
$\bullet$ denotes a BH, $\oplus$ denotes a WD and the number is the mass of the star. 
The eccentricity of the BH-BH binary at the time is also given. 
All listed single BHs are ejected from the cluster after their interaction. 
The two red arrows indicate interactions in which the BH-BH binary recoils out of the 
cluster core. 
}
\label{Fig3}
\end{center}
\end{figure*}

\section{A DYNAMICALLY INFLUENCED BLACK HOLE BINARY MERGER}

As the cluster evolves BH--BH binaries form, generally through three-body encounters 
and exchange interactions. 
Of particular interest is a system that formed at $7.5\,$Gyr comprising BHs of mass 
$9.1$ and $8.2 \, M_\odot$. 
These started life as single main-sequence stars of 
$20$ and $19.5 \, M_\odot$ and quickly evolved to become BHs. 
After $4.8\,$Gyr of evolution the $8.2 \, M_\odot$ BH formed a binary with a 
$1.1 \, M_\odot$ carbon-oxygen WD after a brief three-body 
interaction with the WD and a lower-mass main-sequence star. 
At about $7.5\,$Gyr with the BH--WD binary and the $9.1 \, M_\odot$ BH residing in 
the core of the cluster, a brief interaction ensues and the single BH exchanges 
into the binary at the expense of the WD leaving a BH--BH binary with an orbital period of 
about $6\,800\,$d and an eccentricity of $0.6$. 
Only a few Myr after formation this system is involved in a resonant exchange 
interaction with a third BH of mass $7.0 \, M_\odot$. 
This third BH subsequently leaves the original BH--BH intact but now with an 
orbital period of $410\,$d. 
The $7.0 \, M_\odot$ BH escapes from the cluster with a velocity of $19.5 \, {\rm km} \, {\rm s}^{-1}$. 
The BH--BH binary then remains bound until an age of $11.2\,$Gyr. 
However, in the intervening period it suffers a succession of interactions with other 
stars that harden the system, reducing the orbital period and making the system more 
strongly bound. 
At formation the BH--BH binary resides in the core and successive weak encounters drive the 
period down to $360\,$d until, at around $7.9\,$Gyr, a stronger encounter reduces 
the orbital period to $245\,$d but also causes the binary to recoil to the outer 
regions of the cluster. 
The $6.9 \, M_\odot$ single BH involved in this interaction is ejected from the cluster. 
By $8.9\,$Gyr the BH-BH binary has sunk back into the core and a new encounter, 
with a $11.1 \, M_\odot$ BH and a $1.0 \, M_\odot$ WD, 
reduces the orbital period to $120\,$d. 
Successive encounters over a $2\,$Gyr timeframe reduce the orbital period 
to $6.3\,$d and a stronger encounter then causes a further period decrease 
to $5.5\,$d and once again pushes the binary out of the core. 
This explains the momentary increase in the half-mass radius of the BHs seen  
in Fig.~\ref{Fig2} just before $11\,$Gyr. 
Our BH--BH binary then returns to the core and further encounters reduce the 
orbital period to about $1\,$d (with an eccentricity of 0.6) 
at which point gravitational radiation takes over 
and the system quickly merges to leave a single $17.4 \, M_\odot$ BH at $11.2\,$Gyr 
(under the simple assumption that no mass is lost). 
This is the dynamically influenced BH--BH merge  
-- a gravitational wave source manufactured by the dense cluster environment. 
The evolution of the orbital period for this binary from formation to merging is shown 
in Fig.~\ref{Fig3} with some of the major interactions denoted. 
This highlights that in addition to the interactions which we have described 
there are many minor modifications to the orbital period as the BH--BH binary evolves.

The $17.4 \, M_\odot$ BH quickly captures a companion, a $1.3 \, M_\odot$ 
oxygen-neon WD, in an orbit that reduces from a period of $470\,$d to $80\,$d over 
a $50\,$Myr interval. 
The eccentricity is pumped up to $0.99$ and 
for such high eccentricity the presence of Kozai cycles (Kozai 1962) is inevitable. 
Subsequently the two stars collide at periastron and merge. 
We assume that all of the WD mass is added to the BH to leave a BH of 
mass $18.7 \, M_\odot$. 
This more massive BH immediately captures a new companion and has a 
succession of partners, various WDs and a NS, before settling down with 
an oxygen-neon WD of mass $1.3 \, M_\odot$. 
At an age of $11.7\,$Gyr this binary escapes from the cluster with an eccentricity 
of $0.5$ and an orbital period of $2\,$d. 
This leaves only one BH remaining in the cluster, with a mass of $5.5 \, M_\odot$. 
At an age of $13.9\,$Gyr this collides with an oxygen-neon WD while in a three-body 
system and increases its mass to $6.9 \, M_\odot$. 
It subsequently escapes from the cluster at an age of $14.4\,$Gyr in a binary 
with another oxygen-neon WD and an orbital period of $3.1\,$d. 
The velocity of escape is $32 \, {\rm km} \, {\rm s}^{-1}$ and comes from the break-up 
of a three-body system in which the third star, also an oxygen-neon WD, is ejected 
with a velocity of $210 \, {\rm km} \, {\rm s}^{-1}$.  

As we see in Fig.~\ref{Fig2} the core radius has a local minimum at $11.3\,$Gyr 
followed by a period of expansion until $11.7\,$Gyr when the main core-collapse 
phase resumes. 
This timeframe of $400\,$Myr corresponds with the presence of the $18.7 \, M_\odot$ BH  
in the core as a member of a binary which provides a heating effect through interactions 
with nearby stars. 
This ends when the BH--WD system escapes (as described above). 
The increase in the core radius between $13.9$ and $14.4 \,$Gyr, which temporarily 
halts post-core-collapse oscillations (Breen \& Heggie 2014), 
appears to be similarly the result of heating 
from the BH--WD binary involving the $6.9 \, M_\odot$, which escapes at $14.4\,$Gyr 
(also as described above) and leaves the cluster bereft of BHs.

\begin{figure*}
\begin{center}
\includegraphics[width=\columnwidth]{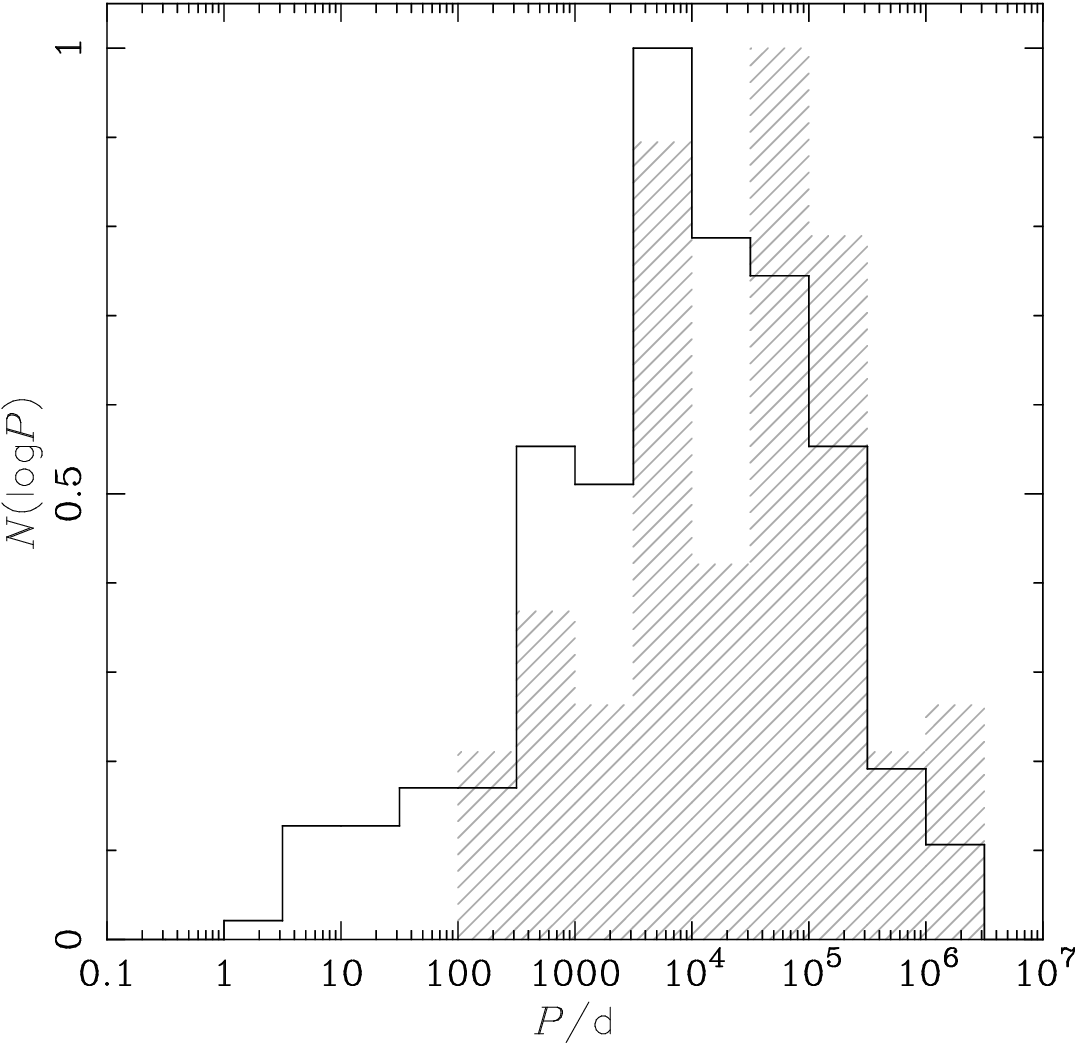}
\caption{Period distribution of BH--BH systems at formation (hatched) and for all distinct 
BH--BH periods that occurred during the simulation (solid line: the {\it modified} distribution). 
Both are normalised so that the maximum number 
of systems in a $\log P$ bin is 1. }
\label{Fig4}
\end{center}
\end{figure*}

\section{BLACK HOLE BINARY PERIOD DISTRIBUTIONS}

There are 84 distinct BH--BH binaries that form during the simulation. 
These are all formed dynamically with exchange interactions being the 
most common pathway. 
None come from primordial binaries even though there are some primordial 
binaries in which both stars are massive enough to evolve to become BHs. 
The mass loss during BH formation or the energy change owing to the 
associated velocity kicks is enough to unbind the binary in each case. 

The period distribution for the 84 BH--BH binaries is shown in Figure~\ref{Fig4}. 
This is the period at formation for each binary. 
The period can subsequently undergo modification, as in the 
examples illustrated in the previous section. 
Thus we also show the distribution of the full range of periods experienced 
by BH--BH binaries during the simulation. 
An individual binary can contribute multiple times to this {\it modified} distribution depending on how often 
its period is changed in an encounter. 
We see that periods of less than $100\,$d are only reached after formation. 
We find that six BH-BH binaries escape from the cluster intact over the course of the evolution. 
These have periods ranging from $220$ to $20\,000\,$d 
and are not expected to merge within a Hubble time. 

For comparison we can look at the lower density model of Sippel \& Hurley (2013) 
which formed 50 distinct BH--BH binaries. 
BH--BH orbital periods less than $1\,000\,$d were not reached in the Sippel \& Hurley (2013) model 
and not surprisingly there were no recorded BH--BH merges in that model. 
That gravitational wave events are more likely to occur in high-density clusters such as 
our current model is to be expected because the timescale for close encounters between stars and binaries 
and for exchange interactions to occur is inversely proportional to the stellar 
number density (Hills 1992). 
In fact Rodriguez, Chatterjee \& Rasio (2016) have shown how the separation, or orbital period, of 
dynamically formed binaries is related to the cluster properties of total mass and half-mass 
radius, with a smaller half-mass radius leading to closer systems which are then 
more likely to merge (Moody \& Sigurdsson 2009).

The Hurley \& Shara (2012) model formed only 
one BH--BH binary and that subsequently escaped intact. 
This simulation used the Kroupa, Tout \& Gilmore (1993) IMF which, 
as noted by Wang et al. (2016), 
is less top-heavy than the Kroupa (2001) IMF, meaning a smaller proportion 
of massive stars and thus fewer BHs. 
Also, the metallicity used was $Z = 0.001$. 
This raises the minimum mass for BH 
production slightly and the lower density of the model means a lower escape velocity 
and thus a decreased retention rate of BHs after velocity kicks. 
All of these factors combined, but primarily the IMF choice, result in only a small number 
of BHs, four in fact, present in the Hurley \& Shara (2012) model at an age of $50\,$Myr 
after comparable velocity kicks to those employed in the current model. 

That all of the BH--BH binaries formed in our model are dynamical in origin 
compares favourably with the $N$-body models made by Ziosi et al. (2014) 
to investigate BH--BH characteristics in young star clusters. 
With models of $N = 5\,000$ and central densities of $10^3 - 10^4 \,$stars ${\rm pc}^{-3}$, 
evolved over a timescale of $100\,$Myr they found that $98\,$per cent or more of the BH--BH binaries 
formed by dynamical exchanges. 
They also found a period distribution primarily populated between $1$ and $10^7 \,$yr 
which allows for much wider systems than we find for our model shown in Fig.~\ref{Fig4}.  


\section{SUMMARY AND DISCUSSION}

So what can we infer from a single instance of a merging binary BH? 
Statistically we do not have a sample that allows us to compute rates, 
as was done by Rodriguez, Chatterjee \& Rasio (2016) from a suite of Monte Carlo 
cluster models ranging in $N$ from $2 \times 10^5$ to $2 \times 10^6$ stars. 
They found a merging rate of about $5 \, {\rm Gpc}^{-3} \, {\rm yr}^{-1}$ 
in the local Universe from dynamically formed BH--BH binaries in GCs and that, 
depending on the velocity kick imparted to BHs at birth, the rate of merging  
from GCs can be comparable to that from field binaries. 
The advantage of the Monte Carlo method is that it allows models to be completed 
on a timescale of days to weeks compared to the approximately six months it took 
for our $N$-body model. 
For a model of similar $N$, initial size and metallicity, 
Rodriguez, Chatterjee \& Rasio (2016) found $3-4$ merges per model. 
In the absence of repeated instances of our $N$-body model and 
considering the stochastic nature of the dynamical formation process, 
our result of one merge can be taken as indicative of an expected $0-2$ per model, 
if we did have the luxury of performing repeated models on a short enough timescale. 
Considering the differences in the modelling approaches, such as the treatment 
of the tidal field (see Giersz et al. 2013), and uncertainties involved with stellar/binary 
evolution (see below), this order of magnitude agreement is pleasing. 

Even though the statistics of dynamically formed BH binary merges have 
been explored previously in Monte Carlo models, the direct confirmation and illustration 
of the process for an individual system in an $N$-body model adds to the picture. 
That has been the intention of this work. 
Going forward 
there is a need to perform a wider range of $N$-body models to look further at how the 
BH--BH binary population characteristics depend on cluster properties, improve statistics 
and allow a more detailed comparison with the Monte Carlo results,  
such as the most recent results of Rodriguez et al. (2016). 
Looking closely at the sister population of BH-NS systems is also warranted. 
In constructing this range of models it would be ideal if $N$ could be varied up to $10^6$ stars. 
Wang et al. (2016) have recently published a set of million-body simulations performed 
with {\tt NBODY6++GPU}, a parallelised version of {\tt NBODY6} that can run across multiple 
GPUs. 
However these models, evolved to $12\,$Gyr, took up to a year to run, were low density 
(about $10^3 \,$stars ${\rm pc}^{-3}$) and did not reach core-collapse. 
A higher density model was evolved to an age of $1\,$Gyr in $120\,$d using 
eight compute nodes with a combined 16 GPUs. 
By comparison the $N \simeq 450\,000$ model of Heggie (2014) with the 
standard non-parallel {\tt NBODY6} was evolved at a higher density but took over 
two years to reach $12\,$Gyr. 
Therefore we are not yet at the stage where we can readily perform a suite of models 
at large $N$. 
In the meantime we shall focus on high-density models in the range of $N = 100\,000$ to $200\,000$ which will each take six or fewer months to complete. 

When composing future models aimed at understanding the BH--BH binary population we 
need to be mindful of the results of Chatterjee, Rodriguez \& Rasio (2016). 
Their work showed that, of the various model assumptions relating to stellar and binary evolution, 
it is the choices for the stellar IMF and the way that post-supernova velocity kicks are applied 
that have the most impact, more so than the initial binary fraction and choices relating to the 
initial binary properties, for example. 
The impact of velocity kicks on BH--BH binary characteristics has also been explored 
recently by Belczynski et al. (2016b). 
Our decision to choose NS and BH velocity kicks from a uniform distribution between 
$0$ and $100 \, {\rm km} \, {\rm s}^{-1}$ is motivated by suggestions that NS retention 
in rich globular clusters is in the range of 10-20 per cent 
(Pfahl, Rappaport \& Podsiadlowski 2002), as achieved by our model,  
and a desire for simplicity in the absence of a definitive kick model. 
For more massive clusters, with larger escape velocities, we would need to increase the 
upper limit of the uniform distribution accordingly so as to maintain retention 
in the 10 to 20 per cent range. 
Conversely, if we were to adopt a Maxwellian distribution with a dispersion of $265 \, {\rm km} \, {\rm s}^{-1}$ 
(Hobbs et al. 2005) in the current model the retention fraction of NSs would be close to zero. 
Similarly for BHs if they were assumed to follow suit. 
Given that NSs and BHs will be the heaviest cluster members and segregate to the centre 
of the cluster their number will affect cluster properties such as the size of the core 
(Breen \& Heggie 2013; Chatterjee, Rodriguez \& Rasio 2016) 
as well as the characteristics of the BH--BH and BH--NS populations. 
As such the manner in which post-supernova velocity kicks are applied deserves serious 
consideration in future models. 

We should also be prepared to relax the upper limit to the stellar IMF and allow initial 
masses up to $100 \, M_\odot$ in future models. 
For the Kroupa (2001) IMF that we have adopted, less than 0.05 per cent of stars lie 
in the $50$ to $100 \, M_\odot$ range, meaning that our decision to cap the IMF at 
$50 \, M_\odot$ has deprived the model of about 50 higher mass stars that would have 
been present if we had extended the IMF to $100 \, M_\odot$. 
This high-end truncation will affect the mass spectrum of the BHs that are produced, 
assuming they are retained in the cluster, and in turn will have an impact on results 
such as the timescale of BH ejection and the cluster evolution 
(e.g. Morscher et al. 2015). 
Increasing the upper limit to the IMF and widening the range of BH masses in the model 
will also increase the applicability to the detection results. 
The LIGO BH--BH binary gravitational wave detection GW150914 (Abbott et al. 2016a) 
has component masses of around $30 \, M_\odot$ and a chirp mass of $28 \, M_\odot$. 
Our BH--BH merger has a lower chirp mass of $7.5 \, M_\odot$ coming from BHs of the order 
of $10 \, M_\odot$. 
It is within the range of the masses derived for the subsequent gravitational wave detection 
GW151226 (Abbott et al. 2016b). 

Within our {\tt NBODY6} treatment the gravitational radiation timescale for close binary systems 
is given by the classical Peters (1964) expression and implemented in terms of changes to the 
orbital angular momentum and the eccentricity (Hurley, Tout \& Pols 2002). 
While this is accurate enough for our current purpose, a more accurate treatment involving 
post-Newtonian terms is available in the sister code {\tt NBODY7} (Aarseth 2012). 
Newtonian dynamics can initiate substantial shrinkage of the orbital separation but 
a favourable timescale for merging is enhanced by high eccentricity which implies that Kozai cycles 
(Kozai 1962) will be present. 
In this case the post-Newtonian treatment would ensure greater accuracy for the modelling 
of the final stages (see Aarseth 2012). 

Our model reaches the end of the main core collapse phase at an age of roughly $12\,$Gyr 
and has only one BH remaining at this late dynamical time. 
Shortly afterwards it has completely lost the BH population. 
This is consistent with predictions from theory and scattering experiments that model  
the behaviour of the BH subsystem that old GCs would retain at most a few BHs 
(e.g. Kulkarni, Hut \& McMillan 1993; Sigurdsson \& Hernquist 1993). 
It is also consistent with the $N$-body model of Heggie (2014) which had lost its BH 
population at about the time that the main core collapse phase ended. 
In contrast the less evolved model of Sippel \& Hurley (2013) had 16 BHs remaining at $12\,$Gyr 
and the million-body models of Wang et al. (2016), also less evolved but with a more generous 
BH retention rate owing to the chosen application of velocity kicks, contained of the order 
of a few hundred to a thousand BHs. 
The set of Monte Carlo models published by Morscher et al. (2015) also showed that 
thousands of BHs could be retained in globular clusters, with some dependence on 
cluster properties and again with a generous retention rate. 
These results highlight the difference between dynamically old and dynamically young clusters in 
terms of expected BH population numbers and the need to further understand how these 
numbers and properties depend on the nature of cluster evolution. 

While our model shows much dynamical activity involving BHs, including the dynamically formed 
gravitational wave source that we have highlighted, what is surprising is that we see little 
evidence for heightened dynamical production of some other exotic binary populations. 
For example, the model contains zero cataclysmic variables and low-mass X-ray 
binaries at $12\,$Gyr. 
Given the higher density of the model compared to earlier simulations and in light of 
the long-held suggestion that a dense stellar environment should produce an increase 
in compact binaries (e.g. Fabian, Pringle \& Rees 1975; Davies 1997; Pooley et al. 2003) these 
non detections are surprising. 
This definitely requires further investigation. 
Most likely the particulars of the primordial binary population, 
for example the choice of a uniform mass-ratio distribution, 
and how this interacts with 
the stellar environment are important for the production of these types of systems, 
but not for BH--BH binaries (Chatterjee, Rodriguez \& Rasio 2016).

\begin{acknowledgements}
JRH would like to thank Churchill College for a Visiting By-Fellowship and the Institute 
of Astronomy visitor program for facilitating this work. 
CAT thanks Churchill College for his fellowship. 
This work was performed on the gSTAR national facility at Swinburne University of Technology. gSTAR is funded by Swinburne and the Australian GovernmentÕs Education Investment Fund. 
\end{acknowledgements}

\end{document}